\documentclass[prb,twocolumn,superscriptaddress]{revtex4}
\usepackage{amsmath}
\usepackage{epsfig}
\usepackage{dcolumn}
\usepackage{bm}
\usepackage{slashed}
\usepackage[toc,page]{appendix}
\usepackage{float}
\usepackage{amsmath}
\usepackage{amssymb}
\usepackage{epsfig}
\usepackage{hyperref}

\usepackage{graphicx}

\def\gapp{\lower.35em\hbox{$\stackrel{\textstyle>}{\sim}$}}
\def\lapp{\lower.35em\hbox{$\stackrel{\textstyle<}{\sim}$}}

\begin{document}
\bibliographystyle{apsrev}

\title{Consequences of a condensed matter realization of Lorentz violating QED in Weyl semi-metals}

\author{Adolfo G. Grushin}

\affiliation{Instituto de Ciencia de Materiales de Madrid, CSIC, Cantoblanco, E-28049 Madrid, Spain.}

\begin{abstract}In Lorentz violating quantum electrodynamics (QED) it is known that a radiatively induced Chern-Simons term appears in the effective action for the gauge field, which is finite but undetermined. This ambiguity is shown to be absent in a condensed matter realization of such a theory in Weyl semi-metals due to the existence of a full microscopic model from which this effective theory emerges. Physically observable consequences such as birefringence are also discussed in this scenario.
\end{abstract}
\maketitle

\section{Introduction}
                          
Historically, bridges between high energy theories and condensed matter phenomena have proven very useful for both communities \cite{VolHel03}, the renormalization group being possibly the hallmark of such a symbiosis \cite{Amit}. More recently, with the advent of low dimensional electronic systems such as graphene, described by a 2+1 dimensional massless Dirac equation \cite{Sem84,VKG10}, there exists the real possibility to explore low dimensional field theories and compare them to actual experiments. For instance, it was predicted that the low energy field theory describing electrons in graphene, being a renormalizable theory, would generate a flow of the Fermi velocity, the only bare parameter of the theory, towards a free theory in the infrared \cite{GGV94}, a fact confirmed by experiment very recently \cite{EGM11}. Other appealing directions regarding graphene include field theories in curved spaces \cite{JCV10} and Schwinger pair production \cite{KV12}.\\
Even more recently, topological field theories have encountered novel physical realizations in materials generically dubbed as topological insulators \cite{QHZ08,HK10,QZ11}, which are bulk insulators that have 2+1 Dirac fermions on the surface. These materials have enabled the theoretical possibility of realizing axion electrodynamics \cite{Wil87} in condensed matter systems and other axion related phenomena \cite{GC11,GRC11}. More importantly, superconducting versions of these materials have been suggested as the root for obtaining an effective realization of the elusive Majorana fermion \cite{Maj37}, which can open new routes towards quantum computation \cite{Been11}.\\
In this work I will put forward an example of Lorentz violating QED that can be realized with a novel class of materials known as Weyl semi-metals \cite{B11,XTVS11,BB11,BHB11}. With the appropriate choice of parameters, these systems of materials can host low energy quasiparticles which are described by the Weyl equations. In the general case however, the low energy quasiparticles are well described by the 3+1 massive Dirac equation. Concretely, as will be shown below, the effective low energy theory resembles a relativistic field theory which can be then modified with appropriate perturbations, so as to take the form of a Lorentz violating version of QED, described by the following action:
\begin{eqnarray}\label{genericLVQED}
S=\int dx^4\bar{\psi}\left(i\slashed{\partial} - m -\slashed{b}\gamma_{5}-e\slashed{A}\right)\psi ,
\end{eqnarray}
with $b_{\mu}$ being a constant four vector. In condensed matter, this is not the first example of such a theory, being ${}^3$He a particularly fruitful example \cite{VolHel03,Vol99,KV05}. \\
In high energy physics, particularly in the context of potential extensions to the standard model, the possibility of a violation of Lorentz symmetries in QED of the form \eqref{genericLVQED} has been subject of intense theoretical research over more than a decade now \cite{CFJ90,CK98,JK99}. Although it seems that our universe is to a very high accuracy Lorentz invariant \cite{CFJ90,KA11}, finding a coherent formulation of Lorentz violating QED seems to be challenging and has generated a very active theoretical debate \cite{ColGlash99,JK99,Chung99,Perez99,Vol99,Chen01,AGS02,NPY07,GNP07,Chen07,CFFP10}. In particular from the beginning it was realized \cite{JK99} that an action of the form \eqref{genericLVQED} generated a Chern Simons term in the effective action for the electromagnetic gauge field of the form $\frac{1}{2}k_{\mu}\tilde{F}^{\mu\nu}A_{\nu}$ where $\tilde{F}^{\mu\nu}=\epsilon^{\mu\nu\rho\sigma}F_{\rho\sigma}$. Intriguingly, the coefficient of this term turns to be finite but undetermined \cite{Jack99}. Ever since, there has been considerable theoretical work in order to clarify this issue under several perspectives. On the one hand Fujikawa type analysis \cite{Chung99}, as well as other non perturbative (in $b$) considerations \cite{Perez99} provided evidence in favour of the ambiguity, whenever the theory was massive \cite{Perez99,GGPW08}. On the other hand, several works have suggested under different symmetry and causality considerations that there is no room for such a correction and that it should vanish \cite{ColGlash99,Bon01,Adam01,CFFP10}. A representative list of possible realizations of $k_{\mu}$ can be found in \cite{Chen01}.\\
In this work, making use of the condensed matter realization of such a theory in the context of Weyl semi-metals, it is shown that a finite and determined value of the radiatively induced Chern Simons term can be fixed unambiguously. In this case, this is possible owing to the fact that a high energy theory exists for this particular system \cite{Vol99}, originated in the microscopic model of the Weyl semi-metal, that enables to determine an unambiguous value of $k_{\mu}$.\\
The paper is structured as follows. In section \ref{Emergence} the exact connection between a low energy description of Weyl semi-metals and Lorentz violating QED will be established. Then, in section \ref{Radiatively} the radiatively induced Chern Simons term will be derived making emphasis on the peculiarities of this particular condensed matter system. In section \ref{Fixing}, the microscopic theory will be reviewed to fix the uncertainty in the low energy theory. Since the Chern Simons term modifies Maxwell's equations inside this material  in section \ref{Physical} some physical implications of this term such as birefringence will be discussed. Finally the main conclusions are presented in section \ref{Conclusions}.
\section{Emergence of Lorentz Violating QED in Weyl semi-metals\label{Emergence}}
\begin{figure}
\includegraphics[scale=0.1]{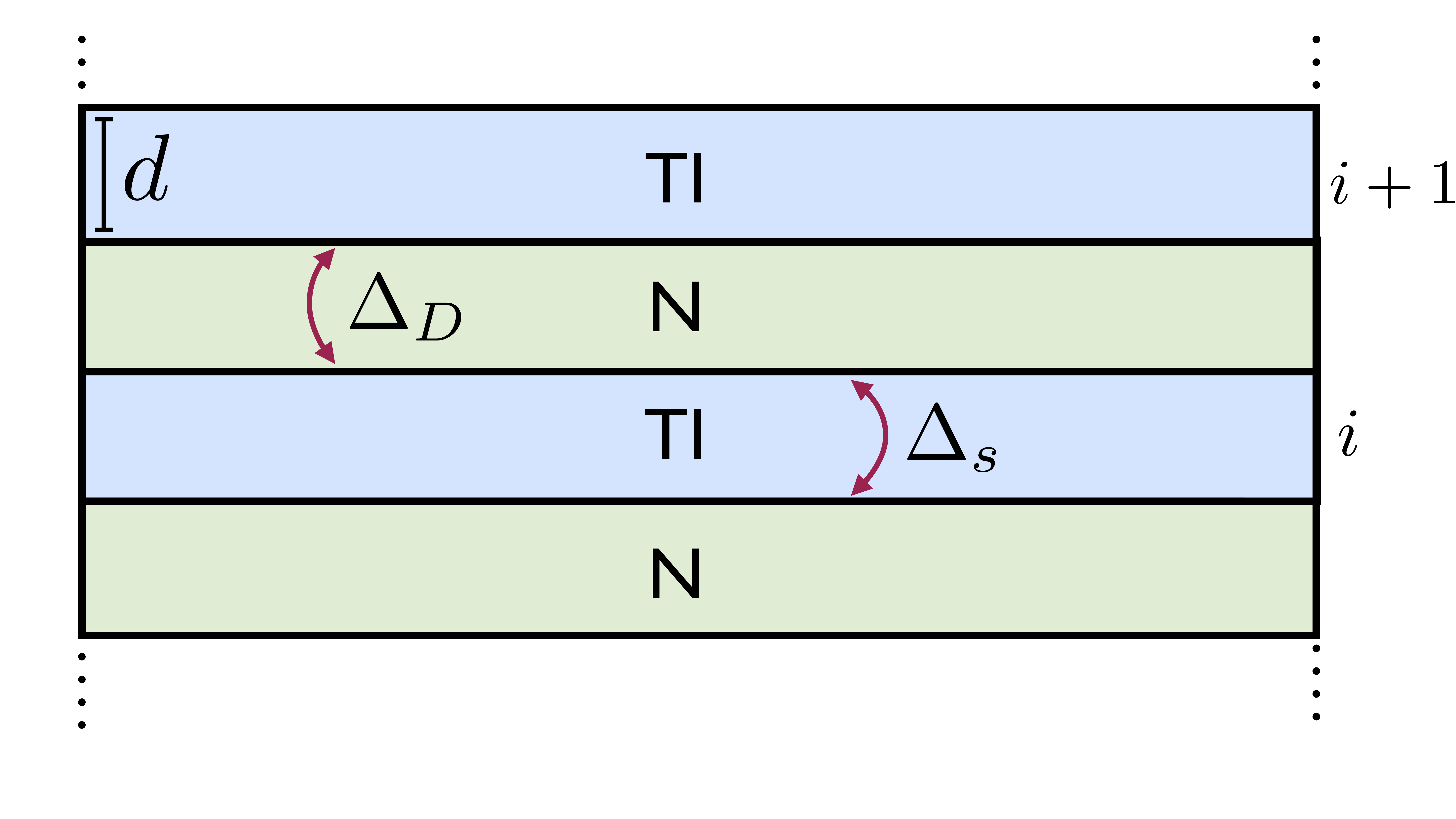}
\caption{\label{Fig:syst}  Illustrative picture of a system that realizes the Weyl semi-metal phase at low energies: a periodic array of alternating normal (N) and topological insulators (TI) \cite{BB11}.  The various parameters of the model are shown schematically: $\Delta_{s}$ is the hopping of an electron to a different surface within the same layer, $\Delta_{D}$ controls the hopping of an electron to a different layer, $d$ is the spacing between topological insulator layers and the symbol $i$ labels the layers.}
\end{figure}
In order to understand how the ambiguity affecting radiative correction is resolved in this context, it is necessary to understand precisely how the action \eqref{genericLVQED} is realized from a microscopic model. This section is thus devoted to provide a pedagogical and self contained introduction to a particular model of Weyl semi-metals \cite{BB11} which will realize the mentioned action with the minimum number of fermionic species.\\
Due to the Nielsen Ninomiya theorem \cite{NielNino81a,NielNino81b}, only an even number of fermions can emerge from a lattice model. To achieve the Weyl semi-metal phase with the minimum number of fermionic species (two), consider, as originally proposed by Burkov and Balents \cite{BB11}, a periodic array of alternating topological insulators and ordinary insulators as shown schematically in Fig. \ref{Fig:syst}. Topological insulators (TI) are 3+1 bulk insulators that posses 2+1 dimensional Dirac fermions at each surface \cite{HK10,QZ11} described by the effective low energy Hamiltonian:
\begin{eqnarray}
H = \sum_{\mathbf{k}_{\perp},i} \left[v_{F} \tau_{z} \otimes(\hat{z}\times \boldsymbol{\sigma})\cdot\mathbf{k_{\perp}}\right]c^{\dagger}_{\mathbf{k_{\perp}}}c_{\mathbf{k_{\perp}}},
\end{eqnarray}
where $\boldsymbol{\sigma}=(\sigma_{x},\sigma_{y})$ represents the spin subspace, $v_{F}$ is the Fermi velocity and $\mathbf{k_{\perp}}=(k_{x},k_{y})$ and $\hat{z}$ is a unitary vector along the growth direction chosen arbitrarily to be in the $z$ direction. The $\tau$ subspace selects at which surface the two species of Dirac fermions live.  The operators $c^{\dagger}_{\mathbf{k_{\perp}}}$ ($c_{\mathbf{k_{\perp}}}$) create (annihilate) quasiparticles at momenta $\mathbf{k_{\perp}}$. Note that the two species of 2+1 Dirac fermions are realized, one at each surface, in concordance to the Nielsen Ninomiya theorem. They can be thought of as the two species or "valleys" realized in graphene, with the pseudospin being here the real spin. \\
When the TI are sufficiently thin, the two surfaces can couple through a hopping amplitude $\Delta_{s}$ which enters the Hamiltonian as: 
\begin{eqnarray}
H_{\Delta_{s}} = \sum_{\mathbf{k}_{\perp},i} \left[\Delta_{s}\tau_{x}\otimes 1_{\sigma}\right]c^{\dagger}_{\mathbf{k_{\perp}}}c_{\mathbf{k_{\perp}}},
\end{eqnarray}
where $1_{\sigma}$ is the identity matrix in spin subspace. In order to couple different surfaces it is necessary to introduce a label $i$ that indicates to which layer the electron is hopping to. If the hopping amplitude is governed by a parameter $\Delta_{D}$ then the coupling between different layers takes the form:
\begin{eqnarray}
H_{\Delta_{D}} = \sum_{\mathbf{k}_{\perp},i} \left[\Delta_{D}\tau^{+}\delta_{i,j+1}+\Delta_{D}\tau^{-}\delta_{i,j-1}
\right]\otimes 1_{\sigma}c^{\dagger}_{\mathbf{k_{\perp}},i}c_{\mathbf{k_{\perp}},j},
\end{eqnarray}
where $\tau^{\pm} \equiv \dfrac{1}{2}\left(\tau_{x}\pm i\tau_{y}\right)$. The full Hamiltonian then reads:
\begin{eqnarray}\label{BalentsHam}
H &=& \sum_{\mathbf{k}_{\perp},i} [v_{F} \tau_{z} \otimes(\hat{z}\times \boldsymbol{\sigma})\cdot\mathbf{k_{\perp}} \delta_{i,j}+\Delta_{s}\tau_{x}\otimes 1_{\sigma}\delta_{i,j}\\
\nonumber
&+&\Delta_{D}\left(\tau^{+}\delta_{i,j+1}+\Delta_{D}\tau^{-}\delta_{i,j-1}\right)\otimes 1_{\sigma}
]c^{\dagger}_{\mathbf{k_{\perp}},i}c_{\mathbf{k_{\perp}},j}.
\end{eqnarray}
Fourier transforming $c^{\dagger}_{\mathbf{k_{\perp}},i}=\sum c^{\dagger}_{\mathbf{k_{\perp}},k_{z}}e^{ik_{z}R_{i}}$, where $R_{m}=dm$ with $m$ an integer and $d$ being the spacing between the layers, the Hamiltonian is diagonalized in momentum space:
\begin{eqnarray}\label{BalentsHamk}
H = \sum_{\mathbf{k}_{\perp},k_{z}} \left[v_{F} \tau_{z}\otimes (\hat{z}\times \boldsymbol{\sigma})\cdot\mathbf{k_{\perp}}+\hat{\Delta}(k_{z}) \right]c^{\dagger}_{\mathbf{k_{\perp}},k_{z}}c_{\mathbf{k_{\perp}},k_{z}},
\end{eqnarray}
where $\hat{\Delta}(k_{z}) \equiv \Delta_{s}\tau_{x}\otimes 1_{\sigma}+\Delta_{D}\left(\tau^{+}e^{ik_{z}d}+\tau^{-}e^{-ik_{z}d}\right)\otimes 1_{\sigma}$. To make the connection with Weyl fermions consider a low energy theory of such a system. The Hamiltonian \eqref{BalentsHamk} has two doubly degenerate eigenvalues given by:
\begin{eqnarray}\label{freebands}
\epsilon^2_{\mathbf{k}} = v_{F}^2(k_{x}^2+k_{y}^2)+\Delta_{s}^2+\Delta_{D}^2+2\Delta_{s}\Delta_{D}\cos(k_{z}d).
\end{eqnarray}
Expanding near $k_{z}=\pi/d$ one obtains:
\begin{eqnarray}
\epsilon^{2}_{\mathbf{k}}&=& v_{F}^2(k_{x}^2+k_{y}^2)+(\Delta_{s}-\Delta_{D})^2+d^2\Delta_{s}\Delta_{D}k_{z}^2,
\end{eqnarray}
where the third momentum is redefined to be $k_{z}-\pi/d\rightarrow k_z$. This Hamiltonian corresponds to a massive 3+1 Dirac fermion at point $\mathbf{k}=(0,0,\pi/d)$ of the Brillouin zone:
\begin{eqnarray}\label{BalentsHamlowk}
H &=& \sum_{\mathbf{k}_{\perp},k_{z}} [v_{F} \tau_{z} (\hat{z}\times \boldsymbol{\sigma})\cdot\mathbf{k_{\perp}}+(\Delta_{s}-\Delta_{D})\tau_{x}\otimes 1_{\sigma}\\
\nonumber
&+&d\sqrt{\Delta_{s}\Delta_{D}}k_{z}\tau_{y}\otimes 1_{\sigma} ]c^{\dagger}_{\mathbf{k_{\perp}},k_{z}}c_{\mathbf{k_{\perp}},k_{z}},
\end{eqnarray}
with dispersion relation:
\begin{eqnarray}
\epsilon_{\pm}(\mathbf{k}) = \pm\sqrt{v_{F}^2(k_{x}^2+k_{y}^2)+\tilde{v}_{F}^2k_{z}^2+m^2},
\end{eqnarray}
where $\tilde{v}_{F}^2=d^2\Delta_{s}\Delta_{D}$ and $m^2=(\Delta_{s}-\Delta_{D})^2$. The Hamiltonian \eqref{BalentsHamlowk} can be recast in a more familiar form:
\begin{eqnarray}\label{DiracW}
H = \sum_{\mathbf{k}} \psi^{\dagger}_{\mathbf{k}}\left( \alpha^{i}k_{i} +\beta m\right)\psi_{\mathbf{k}} ,
\end{eqnarray}
where $i=1,2,3$,  $\psi_{k}=c_{\mathbf{k_{\perp}},k_{z}}$ and the matrices $\alpha$ (following the traditional notation for a Dirac Hamiltonian) are defined as $\alpha_{1}=v_{F}\tau_{z}\otimes\sigma_{y}$, $\alpha_{2}=-v_{F}\tau_{z}\otimes\sigma_{x}$ and $\alpha_{3}=\tilde{v}_{F}\tau_{y}\otimes 1_{\sigma}$ where the Fermi velocities $v_{F}$ are provisionally included inside the definition. A small deviation from $\Delta_{s}/\Delta_{D}=\pm 1$ defines the fourth matrix to be $\beta=\tau_{x}\otimes 1_{\sigma}$. Note in particular that at the critical line $\Delta_{s}/\Delta_{D}=\pm 1$ the quasiparticles are governed by the Weyl equations, where the four component spinor decouples into a pair of two-component Weyl spinors, giving name to the Weyl semi-metal. Away from this critical line, the system is a gapped insulator defined by the Hamiltonian \eqref{DiracW}.\\
With this Hamiltonian in mind, it is possible to write the effective action for the system, which resembles a QED like action, but this time for effective quasiparticles inside the material:
\begin{eqnarray}\label{3+1Dirac}
S = \int \dfrac{d^4k}{(2\pi)^4} \bar{\psi}_{k}\left(\gamma_{\mu}M_{\hspace{2mm} \nu}^{\mu}k^{\nu}-m\right)\psi_{k},
\end{eqnarray}
where it is convenient to introduce the diagonal matrix $M^{\mu}_{\hspace{2mm}  \nu}=diag(1,v_{F},v_{F},\tilde{v}_{F})$ to adequately manage the anisotropic Fermi velocities. The exact relation between $\gamma$ and $\alpha$ and $\beta$ can be obtained with the usual procedure, and it is detailed in appendix \ref{matrices}. \\
In view of the action \eqref{3+1Dirac}, and postponing for the end of this section the discussion concerning Lorentz invariance, a natural question arises: is it possible to introduce a term of the form $\bar{\psi}\slashed{b}\gamma_{5}\psi$ to reproduce an action that resembles \eqref{genericLVQED}? The answer, is indeed affirmative. Doping these materials with magnetic impurities breaks time reversal symmetry \cite{BB11} and introduces a term of the form:
\begin{eqnarray}
H_{m_{1c}} = \sum_{\mathbf{k}}  m_{1c}\psi^{\dagger}_{\mathbf{k}}1_{\tau}\otimes\sigma_{z} \psi_{\mathbf{k}},
\end{eqnarray}
which in the $\gamma$ matrix representation is nothing but $\bar{\psi}\gamma^{3}b_{3}\gamma_{5}\psi$ (see appendix \ref{matrices} for details).
The presence of this term opens a gap of size $m_{1c}$ at the surface of each TI layer. Physically, this term can be understood as arising from a magnetization determined by the density of magnetic impurities. For sufficiently weak magnetizations, the magnetic field only couples to the surface states as a Zeeman field, which for the massless 2+1 Dirac fermions at the surface is exactly a mass term that opens a gap at the surface, being the size of the $m_{1c}$ proportional to the magnetization. Experimentally, it was confirmed in Ref \cite{Chen10} that for the topological insulator Bi$_2$Se$_3$, the magnitude of the gap increases with the impurity density and can be as large as $60$meV for a concentration of $0.12$ Fe impurities per Bi atom.\\
Similarly, in \cite{ZWB12} it was shown that if an inversion breaking spin orbit coupling term is allowed, it has the form:
\begin{eqnarray}
H_{\lambda} = \sum_{\mathbf{k}} \lambda\psi^{\dagger}_{\mathbf{k}}\tau_{y}\otimes\sigma_{z}\psi_{\mathbf{k}}. 
\end{eqnarray}
In this case, this corresponds exactly to a term of the form $\bar{\psi}\gamma^{0}b_{0}\gamma_{5}\psi$. Unfortunately, the precise value of $b_{0}=\lambda$ in this system is still unknown. However, it can still be expected to be large since all TI have intrinsically a large spin orbit coupling which is precisely a necessary ingredient for their topological nature. Note as well that $\lambda$ in this model is assumed to be an independent parameter and therefore can be either larger or smaller than $m$. \\
With these two terms we finally arrive to the action:
\begin{eqnarray}\label{genericLVQEDb}
S=\int \dfrac{dk^4}{(2\pi)^4}\bar{\psi}\left(\gamma_{\mu}M_{\hspace{2mm} \nu}^{\mu}k^{\nu} - m -\slashed{b}\gamma_{5}\right)\psi.
\end{eqnarray}
When coupled to an electromagnetic gauge field, this action is the condensed matter analogue of a Lorentz violating QED with a CPT violating term given by \eqref{genericLVQED} \cite{CK98,JK99}, and constitutes both the starting point and the first result of this work. \\
It is evident that this action, arising from a condensed matter system, breaks Lorentz invariance even without the term proportional to $b$, due to the appearance of the matrix $M_{\hspace{2mm} \nu}^{\mu}$ in the action, which is in a way a trivial violation of the symmetry. As it will be shown below, the consequences of the CPT violating term $\bar{\psi}\slashed{b}\gamma_{5}\psi$ are much more profound, and so they will be the focus of the following sections.

\section{Radiatively induced Chern-Simons term \label{Radiatively}}

As in ordinary QED, the coupling to the external gauge field is determined through a term of the form $j^{\mu}A_{\mu}$, where $j_{\mu}$ is the current operator, defined by the free fermionic action. In this case the current operator is defined in terms of $M^{\mu}_{\hspace{2mm} \nu}$ containing the Fermi velocities:
\begin{eqnarray}
j^{\mu}=M^{\mu}_{\hspace{2mm}\alpha}\bar{\psi}_{\mathbf{k}}\gamma^{\alpha}\psi_{\mathbf{k}}.
\end{eqnarray}
Consider now the quantum expectation value for such a current operator. To one loop, it is defined by the polarization of the photon $\Pi^{\mu\nu}$ and it is given by
\begin{eqnarray}\label{current}\nonumber
\left\langle j^{\mu}\right\rangle  &=& \left\langle M^{\mu}_{\hspace{2mm}\alpha}M^{\nu}_{\hspace{2mm}\beta}\bar{\psi}_{\mathbf{k}}\gamma^{\alpha}\psi_{\mathbf{k}}
\bar{\psi}_{\mathbf{k}}\gamma^{\beta}\psi_{\mathbf{k}}\right\rangle A_{\nu}\\
&=&M^{\mu}_{\hspace{2mm}\alpha}M^{\nu}_{\hspace{2mm}\beta}\Pi^{\alpha\beta} A_{\nu}.
\end{eqnarray}
In terms of Feynman diagrams, $\Pi^{\alpha\beta}$ is the analogous to the QED photon bubble with the only difference that the fermionic propagator is $G(k,b)$ given by:
\begin{eqnarray}\label{Gb}
G(k,b)=\dfrac{i}{\slashed{k} -m - \slashed{b} \gamma_{5}}.
\end{eqnarray}
The integral determining $\Pi^{\mu\nu}$ is given by an appropriate generalization of the one loop vacuum polarization diagram:
\begin{eqnarray}\label{bubble}
\Pi^{\mu\nu}= \dfrac{e^2}{v_{F}^2\tilde{v}_{F}}\int \dfrac{dk^4}{(2\pi)^4}\mathrm{Tr}\left\lbrace\gamma^{\mu}G(k,b)\gamma^{\nu}G(k+p',b)\right\rbrace.
\end{eqnarray}
As before, the prefactor $\frac{1}{v_{F}^2\tilde{v}_{F}}$ stems from a rescaling of the momenta with the corresponding Fermi velocities and $p'^{\mu}=M^{\mu}_{\hspace{2mm} \nu}p^{\nu}$ is the rescaled external four-momentum vector. Once established this connection between Lorentz violating QED and Weyl semi-metals it is straightforward to calculate the odd part of \eqref{bubble} non perturbatively in $b$ following for instance Ref. \cite{Perez99}:
\begin{eqnarray}\label{zerofreq}
\Pi^{\mu\nu}_{\mathrm{odd}}= \dfrac{e^2}{v_{F}^2\tilde{v}_{F}}\epsilon^{\mu\nu\rho\sigma}p^{'}_{\rho}b_{\sigma}\left\{
	\begin{array}{ll}
		 C &\mbox{if } -b^2 \leq m^2 \\
	    C-\dfrac{1}{2\pi^2}\sqrt{1-\dfrac{m^2}{b^2}}  &\mbox{if }  -b^2 \geq m^2
	\end{array}
\right. ,
\end{eqnarray} 
where $C$ is an finite but undetermined constant \cite{JK99,Chung99,Perez99}.  Introducing \eqref{zerofreq} into \eqref{current} one obtains the response of the Weyl semi-metal to an external electromagnetic field in the presence of both spin orbit coupling, given by $b_{0}$ and magnetic impurities governed by $b_{3}$:
\begin{widetext}
\begin{eqnarray}\label{zerofreqcurr}
j^{\mu}_{\mathrm{odd}} = M^{\mu}_{\hspace{2mm}\alpha}M^{\nu}_{\hspace{2mm}\beta}\dfrac{e^2}{v_{F}^2\tilde{v}_{F}}A_{\nu}\epsilon^{\alpha\beta\rho\sigma}p^{'}_{\rho}b_{\sigma}\left\{
	\begin{array}{ll}
		 C &\mbox{if } -b^2 \leq m^2 \\
		 C-\dfrac{1}{2\pi^2}\sqrt{1-\dfrac{m^2}{b^2}} &\mbox{if }  -b^2 \geq m^2
	\end{array}
\right. ,
\end{eqnarray} 
\end{widetext}
This is itself a novel result in the context of Weyl semi-metals being a non perturbative calculation in both the spin orbit coupling and the magnetic impurity strength. However, it is necessary to fix the constant $C$ in order to argue that this is the physical response of the system. This issue is addressed in detail in the next section.

\section{Fixing the ambiguity \label{Fixing}}

As introduced in the first section and argued by many preceding works \cite{ColGlash99,JK99,Chung99,Perez99,Vol99,Chen01,AGS02,Chen07,CFFP10} under very different approaches, the constant $C$ is finite and undetermined. It depends strongly on the regularization method used. It is only in the massless case, that this constant is fixed unambiguously \cite{Perez99,GGPW08}. This ambiguity in the context of Weyl semi-metals seems at least paradoxical, since this constant defines physically measurable observables such as the conductivity, which I proceed to discuss.\\
Consider for example a constant electric field in the $y$ direction. The Hall conductivity, is defined as the off diagonal part of the proportionality tensor between the current and the electric field:
\begin{eqnarray}
j^{x}=\sigma^{xy}E_{y},
\end{eqnarray}
and can be measured in transport experiments.  If $b=(0,0,0,b_{3})$ setting $\mu=x$ in \eqref{zerofreqcurr} gives:
\begin{eqnarray}\label{zerofreqcurrb3}
j^{x}_{\mathrm{odd}}  &=& \dfrac{e^2}{\tilde{v}_{F}}E_{y}b_{3}\left\{
	\begin{array}{ll}
		 C &\mbox{if } b_{3}^2 \leq m^2 \\
		 C-\dfrac{1}{2\pi^2}\sqrt{1-\dfrac{m^2}{b_{3}^2}} &\mbox{if }  b_{3}^2 \geq m^2
	\end{array}
\right. \\
\nonumber
&\equiv& \sigma^{xy}E_{y} ,
\end{eqnarray} 
Does this result imply that the Hall conductivity is ambiguous in these materials? The answer turns out to be negative owing to the fact that there is a microscopic model, or in other words, a high energy theory, from which this conductivity can also be calculated. As it will now be shown, a consistent matching of these two theories will imply that $C=0$.\\
As a starting point consider the limit of decoupled layers where $\Delta_{s}\gg\Delta_{D}$. In this case, the Hall conductivity is determined by the Hall conductivity of the two massive 2+1 Dirac fermions at each surface of the topological insulator and it is proportional to the sum of the signs of the masses of each fermionic species \cite{Sem84}:
\begin{eqnarray}
\sigma_{2D}^{xy}=\dfrac{e^2}{h}\sum_{i} \mathrm{sign}(m_{i}),
\end{eqnarray}
where the sum $i$ runs over all fermionic species, in this case two. In this limit, Hamiltonian \eqref{BalentsHamk} together with the $b_{3}$ perturbation is independent of $k_{z}$ and it is that of two 2+1 massive Dirac fermions with masses $m_{2D}=b_{3}\pm\Delta_{s}$. Whenever $b_{3}<\Delta_{s}$ is satisfied $\sigma_{2D}^{xy}$ vanishes and so does $\sigma_{xy}$. Comparing this result with the corresponding case $b_{3}^2<m^2$ in \eqref{zerofreqcurrb3} one is forced to set: 
\begin{equation}\label{C=0}
C\Big|_\text{WSM}=0.
\end{equation}
This is the central result of this work since this constant is precisely the one that enters the odd part of the polarization tensor \eqref{zerofreq} that determines the response of Weyl semi-metals to an external electromagnetic field.\\
A second argument, perhaps physically more transparent, comes by introducing a finite $\Delta_{D}$. In this case, there is a set of $2+1$ massive Dirac hamiltonians with masses which depend parametrically with $k_{z}$. Thus, only at certain values of the vector $k_{z}$ the Hall conductivity vanishes. The 2D Hall conductivity is now a step function \cite{BB11,BHB11}:
\begin{eqnarray}
\sigma_{2D}^{xy}=\dfrac{e^2}{h}\Theta(k_{0}-|\pi/d-k_{z}|),
\end{eqnarray}
where $k_{0}$ is the separation in $k$ space between the two Dirac fermions given by:
\begin{eqnarray}\label{k0}
k_0 =\dfrac{1}{d}\arccos\left[1-\dfrac{b_{3}^2-m^2}{2\Delta_{s}\Delta_{D}}\right].
\end{eqnarray}
The integral over the whole Brillouin zone defines de conductivity through:
\begin{eqnarray}
\sigma^{xy}=\int_{-\pi/d}^{\pi/d} \dfrac{dk_{z}}{2\pi}\sigma^{xy}_{2D}(k_{z})= \dfrac{e^2k_{0}}{\pi h}.
\end{eqnarray}
The conductivity is proportional to the separation of the Dirac Fermions in the reciprocal space, a result first proven in \cite{BB11}. To compare with \eqref{zerofreqcurrb3} it is necessary to expand \eqref{k0} near $b_{3}^2\sim m^2$ which is the case of the low energy theory. This gives a Hall conductivity of:
\begin{eqnarray}
\sigma^{xy}= \dfrac{e^2k_{0}}{\pi h}\simeq \dfrac{e^2}{\pi h}\dfrac{b_{3}}{d\sqrt{\Delta_{D}\Delta_{s}}}\sqrt{1-\dfrac{m^2}{b_{3}^2}}.
\end{eqnarray}
Restoring $\hbar$ and using that $\tilde{v_{F}}=d\sqrt{\Delta_{D}\Delta_{s}}$ it is straightforward to see that \eqref{zerofreqcurrb3} reproduces this result only if $C=0$ in agreement with \eqref{C=0}. The microscopic theory, which considered the whole Brillouin zone in the $k_{z}$ direction has fixed the value of the arbitrary but finite constant generated in the low energy theory to be zero. As it turns out, having a \emph{lattice} model from which the low energy theory is derived regularizes the theory to fix the ambiguity. \\
A final comment is in order. With this analysis it has been proven that a zero Hall conductivity for the lattice model would always imply a zero Hall conductivity calculated within the low energy effective field theory and fixes $C=0$. However, the inverse statement is not always true. This means that, if one calculates a non zero Hall conductivity in the lattice model, it might be proportional to a lattice vector $\mathbf{G}_{j}$ \cite{KHW93}, a result which the effective low energy field theory approach will never obtain. One can check with a similar analysis to the one described above that this is indeed the case for Hamiltonian \eqref{BalentsHamk} when $b_{3} \gtrsim\Delta_{s}+\Delta_{D}$ \cite{BB11}.

\section{Physical consequences \label{Physical}}

I now turn to discuss measurable physical consequences derived from this theory. As it is well known, the complete polarizability $\Pi^{\mu\nu}$ modifies Maxwell's equations inside the material and will govern the electrodynamic response of this system. Its even part $\Pi^{\mu\nu}_{\mathrm{e}}$ will define the dielectric function and the magnetic permeability of the material while its odd part will add novel terms which will drastically modify the response of the material to an external perturbation. Integrating out fermions the effective action for the gauge field inside the material is:
\begin{eqnarray}\label{effecaction}
S = \int \dfrac{d^4k}{(2\pi)^4} A_{\mu}\Pi_{\mathrm{e}}^{\mu\nu}A_{\nu}-\dfrac{1}{2}s_{\mu}\tilde{F}^{\mu\nu}A_{\nu},
\end{eqnarray}
where $s^{\mu} = \dfrac{e^2}{v_{F}^2\tilde{v}_{F}}\sqrt{1-\dfrac{m^2}{b^2}}b^{\mu}$ is defined by \eqref{zerofreq} whenever $-b^2\geq m^2$ and zero otherwise and $\tilde{F}^{\mu\nu}=\epsilon^{\mu\nu\rho\sigma}k'_{\rho}A_{\nu}$. This action is known to generate a modified version of Maxwell's equation \cite{CFJ90}, although in this case it is necessary to keep track of the anisotropic Fermi velocity. For the case where only $b_{3}$ is non zero it is possible to write a clean set of Maxwell equations given by:
\begin{eqnarray}\label{ME1}
\nabla\cdot \mathbf{D} &=& 4\pi\rho - v_{F}^2\mathbf{s}\cdot \mathbf{B},\\
\label{ME2}
\nabla \cdot \mathbf{B}&=&0,\\
\label{ME3}
\nabla\times \mathbf{E}&=&-\dfrac{1}{c}\dfrac{\partial\mathbf{B}}{\partial t},\\
\label{ME4}
\nabla\times\mathbf{H} &=& \mathbf{j}+ \dfrac{\partial\mathbf{D}}{\partial t} +v_{F}^2\mathbf{s}\times \mathbf{E},
\end{eqnarray}
where now $s_{\mu}=(0,0,0,s_{3})$ and the $v_{F}^2$ coefficient is fixed by \eqref{zerofreqcurrb3}. Note the similarity between these equations and the ones corresponding to axion electrodynamics \cite{Wil87} which could be realized in topological insulators \cite{QHZ08}. As in topological insulators, the novel terms in the equations of motions give rise to new physical phenomena.\\
Consider as an example, light propagation in a Weyl semi-metal system described by \eqref{ME1}-\eqref{ME4}. From the source free $(\rho = 0,\mathbf{j}=0)$ equations it is easy to derive the following wave equation inside the Weyl semi-metal:
\begin{eqnarray}
\nabla^2 E-\dfrac{1}{c_{w}^2}\dfrac{\partial^2 E}{\partial t^2}+\nabla\left(\nabla\cdot E\right)= v_{F}^2\mathbf{s}\times\dfrac{\partial E}{\partial t},
\end{eqnarray}
where $\mu\sim 1$ is assumed being satisfied in a wide range of frequencies \cite{LLv8} and $c_{w}=1/\sqrt{\epsilon}$ is the velocity of light inside Weyl semi-metals. Following \cite{CFJ90}, it is possible to derive the dispersion relation that photons entering the Weyl semi-metal should satisfy: 
\begin{eqnarray}
\left(\dfrac{\omega^2}{c_{w}^2}-k^2\right)^2-v_{F}^2(\omega^2- k^2)s_{3}^2=(v_{F}^2k_{3} s_{3})^2,
\end{eqnarray}
which characterize a birefringent media that in this approximation is due entirely to the induced Chern-Simons term. Birrefrengence of this kind will be generic to all Weyl semi-metals phases that posses a term of the form $\bar{\psi}\slashed{b}\gamma_{5}\psi$, independent of the particular microscopic model that realizes such a phase, an in particular, on the extrinsic details of the lattice model. An observation of birefringence with light of sufficiently long wavelength would provide an experimental measurement of the constant $C$. For example, linearly polarized light entering such a medium will leave it in the form of elliptically polarized light. \\
The action \eqref{effecaction} can potentially host other interesting physical phenomena. One very appealing possibility is that these materials, in analogy with topological insulators \cite{GC11}, could posses a repulsive Casimir effect which might be suppressed but still exist in anisotropic materials like the ones described here \cite{GRC11}. \\
Finally, the coupling between magnetic and electric degrees of freedom can enable routes towards exploring exotic phenomena similar to the effective magnetic monopoles possible in axion electrodynamics \cite{QRJZ09}  and chiral gauge fields \cite{CPQ12}.

\section{Conclusions \label{Conclusions}}

In this work the emergence of a Lorentz violating QED in the novel class of materials known as Weyl semi-metals was explored in detail. It was found that Weyl semi-metals in the presence of magnetic impurities and spin orbit coupling realize a Lorentz violating version of QED with a term of the form $\bar{\psi}\slashed{b}\gamma_{5}\psi$. The electromagnetic response of such a system includes a radiatively induced Chern Simons term in the effective action for the electromagnetic gauge field of the form $\frac{1}{2}k_{\mu}\tilde{F}^{\mu\nu}A_{\nu}$ defined by the odd part of the photon self energy $\Pi^{\mu\nu}_{\mathrm{odd}}$. This correction to the photon self energy is finite but undetermined, also in the low energy theory of Weyl semi-metals. However, it has been shown that in this system it is possible to fix the ambiguity due to the existence of a microscopic model from which this result can be derived. The comparison between these two approaches fixes the value for the constant that parametrizes the uncertainty ($C$) to zero for this system.\\
Although in this case, the finite value of $C$ turns to be zero, it is in principle possible that other Weyl semi-metals might realize other values. In particular, the most favourable situation would be that where a Weyl semi-metal phase is realized on the border of an anomalous Hall phase so that there is a finite  Hall conductivity on both sides of such a transition described by an equation of the form of \eqref{zerofreq}. This situation is realized in this model although here, the anomalous Hall conductivity is proportional to a lattice vector \cite{BB11} and will never appear in a low energy description such as the one presented in this work. The analysis presented here can be applied to other more sophisticated examples of Weyl semi-metal systems \cite{XTVS11,G11,YLR11,XWW11,J12,DLC12} to determine whether $C= 0$ is a generic feature in condensed matter systems as suggested by topological arguments \cite{KV05}. \\
In addition to the finite and unambiguous result determined by \eqref{zerofreq} with $C=0$, the formulation in terms of a Lorentz violating QED of the low energy theory of Weyl semi-metals enables to perform calculations non perturbatively both in the spin orbit coupling and the magnetic impurity strength.\\
Finally, it has been shown that when $ -b^2 \geq m$ the coefficient $k^{\mu}$ enters the effective Maxwell equations inside the material, substantially modifying the electrodynamics of the system. In particular the new terms proportional to $b_{\mu}$ will give rise to birefringence when light enters the material. Being a condensed matter system, an observation of such a birefringence is a feasible experiment, in contrast to conventional astrophysical observations which strongly constrain the observability of Lorentz violating QED phenomena. More exotic scenarios derived from the modified Maxwell equations, such as the stationary magnetic order proposed in \cite{CFJ90}, could be realized in these types of systems, although they will only occur in the situation where $C\neq 0$ which will produce the right form of electromagnetic solutions.

\section{Acknowledgments}

I am indebted to M. A. H. Vozmediano for support and encouragment during the completion of this work. I acknowledge stimulating conversations with A. Cortijo, F. de Juan, M. Sturla, H. Ochoa, B. Amorim, P. Goswami and A. A. Burkov. Hospitality of the Kavli Institute of Theoretical Physics and support from the Spanish national projects FIS2011- 23713 and FIS2008-00124 and Brazilian PIB2010BZ-00512 are greatly acknowledged.  


\hspace*{2cm}

\appendix

\section{Matrix definitions:\label{matrices}}
In this appendix, all the definitions for the matrices used in the main text are reviewed. The Hamiltonian matrices are defined to be:

\begin{eqnarray}\nonumber
\alpha_{1}&=&\tau_{z}\otimes\sigma_{y}\\
\nonumber
\alpha_{2}&=&-\tau_{z}\otimes\sigma_{x}\\
\nonumber
\alpha_{3}&=&\tau_{y}\otimes 1_{\sigma}\\
\nonumber
\beta &=&\tau_{x}\otimes 1_{\sigma}\\
\nonumber
\alpha_{5}&=&-\tau_{z}\otimes\sigma_{z}=\beta\alpha_{1}\alpha_{2}\alpha_{3}.
\end{eqnarray}

To construct a low energy effective field theory action the following dictionary can be used, following the usual convention for the $\gamma$ matrices.

\begin{eqnarray}\label{hamall}\nonumber
H_{0}(\mathbf{k}) &=& \psi^{\dagger}_{\mathbf{k}}\alpha^{i}k_{i} \psi_{\mathbf{k}} = \bar{\psi}_{\mathbf{k}} \beta\alpha^{i}k_{i} \psi_{\mathbf{k}} \equiv \bar{\psi}_{\mathbf{k}} \gamma^{i}k_{i} \psi_{\mathbf{k}}\\
\nonumber
H_{m}(\mathbf{k})  &=& m\psi^{\dagger}_{\mathbf{k}}\beta\psi_{\mathbf{k}} \equiv  m \bar{\psi}_{\mathbf{k}}\psi_{\mathbf{k}}\\
\nonumber
H_{m_{1c}}(\mathbf{k})  &=& m_{1c}\psi^{\dagger}_{\mathbf{k}}1_{\tau}\otimes\sigma_{z}\psi_{\mathbf{k}} \\
\nonumber
&=& m_{1c} \psi^{\dagger}_{\mathbf{k}}i\beta\alpha_{3}\alpha_{5}\psi_{\mathbf{k}}\equiv b_{3} \bar{\psi}_{\mathbf{k}}\gamma_{3}\gamma_{5}\psi_{\mathbf{k}}\\
\nonumber
H_{\lambda}(\mathbf{k})  &=& \lambda \psi^{\dagger}_{\mathbf{k}}\tau_{y}\otimes\sigma_{z}\psi_{\mathbf{k}}\\
\nonumber
&=& -\lambda \psi^{\dagger}_{\mathbf{k}}i\beta\alpha_{5}\psi_{\mathbf{k}}\equiv -b_{0}\bar{\psi}_{\mathbf{k}}\gamma_{0}\gamma_{5}\psi_{\mathbf{k}}
\end{eqnarray}

which define $\gamma_{i}=\beta\alpha_{i}$, $\gamma_{0}=\beta$ and $\gamma_{5}=-i\beta\alpha_{5}$, $b_{3}=m_{c1}$ and $b_{0}=\lambda$.

\end{document}